# Effective viscosity of methyl cellulose solutions in phosphate buffered saline in real-time deformability cytometry

Felix Reichel and Beyza Büyükurgancı

Max Planck Institute for the Science of Light and Max-Planck-Zentrum für Physik und Medizin, Erlangen, Germany

**\*Correspondence:**
felix.reichel@mpl.mpg.de

## Abstract

Here, we derive the equations to calculate the effective viscosity of solutions of methyl cellulose (MC) dissolved in phosphate buffered saline (PBS) in real-time deformability cytometry (RT-DC) experiments. The calculations are based on the rheometer measurements described by Büyükurgancı et al. (2023). We outline how to get the final equations and compare the results to the current viscosity model from Herold (2017). These viscosity functions will be used to determine the Young's moduli of biological cells and other soft materials from RT-DC experiments.

## 1  Background

### 1.1  Rheology of MC-PBS solutions

Büyükurgancı et al. used three different rheometer types to measure the viscosity curves of three MC-PBS (0.49 w%, 0.59 w%, 0.83 w%)[*] solutions, commonly used in RT-DC experiments, over a wide range of shear rates ($0.1 - 150{,}000 \text{ s}^{-1}$).[1] They found that at shear rates over $5{,}000 \text{ s}^{-1}$, the viscosity curves can be described with a power law:

$$\eta = K \cdot \left(\frac{\dot{\gamma}}{\dot{\gamma}_0}\right)^{n-1}, \qquad (1)$$

where $\dot{\gamma}$ is the shear rate, $\dot{\gamma}_0$ is a normalization shear rate, which will be set to $\dot{\gamma}_0 = 1 \text{ s}^{-1}$ for convenience, $K$ is the *flow consistency index* and $n$ is the *flow behavior index*. RT-DC measurements are usually performed at shear rates much higher than that, so the power law approximation is a valid approach for RT-DC.

Büyükurgancı et al. investigated the temperature dependence of these solutions and found the following relations for the parameters $K$ and $n$:

$$n = \alpha \cdot T + \beta \qquad (2)$$

---

[*] 0.49 w% is also commonly known as 0.5% MC-PBS, 0.49% MC-PBS or CellCarrier; 0.59 w% as 0.6% MC-PBS, 0.59% MC-PBS or CellCarrier B.

$$K = A \cdot e^{\frac{\lambda}{T}} \tag{3}$$

where $\alpha$, $\beta$, $A$ and $\lambda$ are empirical parameters that were determined by fitting $K$ and $n$ over the temperature $T$ in a range of 22 – 37 °C. The resulting parameters are given in Table 1:

With the values from Table 1 and equations 1-3, one can fully describe the viscosity curve of each MC-PBS solution for shear rates greater $5{,}000 \text{ s}^{-1}$ at a given temperature.

**Table 1: Parameters for temperature dependence of the flow consistency index $K$ and flow behavior index $n$ as reported in Büyükurgancı et al.**

| Solution | $\alpha$ [1/K] | $\beta$ | $A$ [Pa s] | $\lambda$ [K] |
|---|---|---|---|---|
| 0.49% MC-PBS | 0.0022 ± 0.0004 | 0.01 ± 0.11 | $0.8 \cdot 10^{-6}$ ± $1.0 \cdot 10^{-6}$ | 3691.8 ± 475.2 |
| 0.59% MC-PBS | 0.0024 ± 0.0003 | -0.14 ± 0.08 | $1.5 \cdot 10^{-5}$ ± $1.2 \cdot 10^{-5}$ | 3095.6 ± 242.5 |
| 0.83% MC-PBS | 0.0021 ± 0.0003 | -0.12 ± 0.08 | $1.8 \cdot 10^{-5}$ ± $2.5 \cdot 10^{-5}$ | 3351.6 ± 412.4 |

## 1.2 Shear rates and effective viscosity of shear thinning liquids in microfluidic channels

In pressure-driven flows, such as in microfluidic channels, the flow will not have a constant velocity but a distribution with the fastest flow along the centerline of the channel and zero velocity at the channel walls. In the case of a power law liquid, the flow behavior index influences this velocity profile. The velocity profile of a power liquid in a circular channel (pipe flow) can be expressed with the following formula:

$$v(r) = v_0 \cdot \left(1 - \left(\frac{r}{R}\right)^{\frac{n+1}{n}}\right), \tag{4}$$

with the centerline velocity $v_0$ and the channel radius $R$. Setting $n = 1$ yields the equation for a Newtonian liquid. The centerline velocity $v_0$ will be different for Newtonian and shear thinning media. Usually, it is relatively simple to calculate the centerline velocity for a Newtonian liquid from the pressure drop or flow rate and the channel radius (Hagen-Poiseuille law) but it is complicated to calculate for power law liquids. By setting that the volume flow for Newtonian and power law liquids needs to be preserved, one can calculate the centerline velocity of the power law liquid from the centerline velocity of the Newtonian liquid:

$$\int_0^R v_{\text{Newtonian}}(r) \, dr \equiv \int_0^R v_{\text{Power law}}(r) \, dr \tag{5}$$

$$v_{0,\text{Power law}} = v_{0,\text{Newtonian}} \frac{2}{3}\left(\frac{2n+1}{n+1}\right) \tag{6}$$

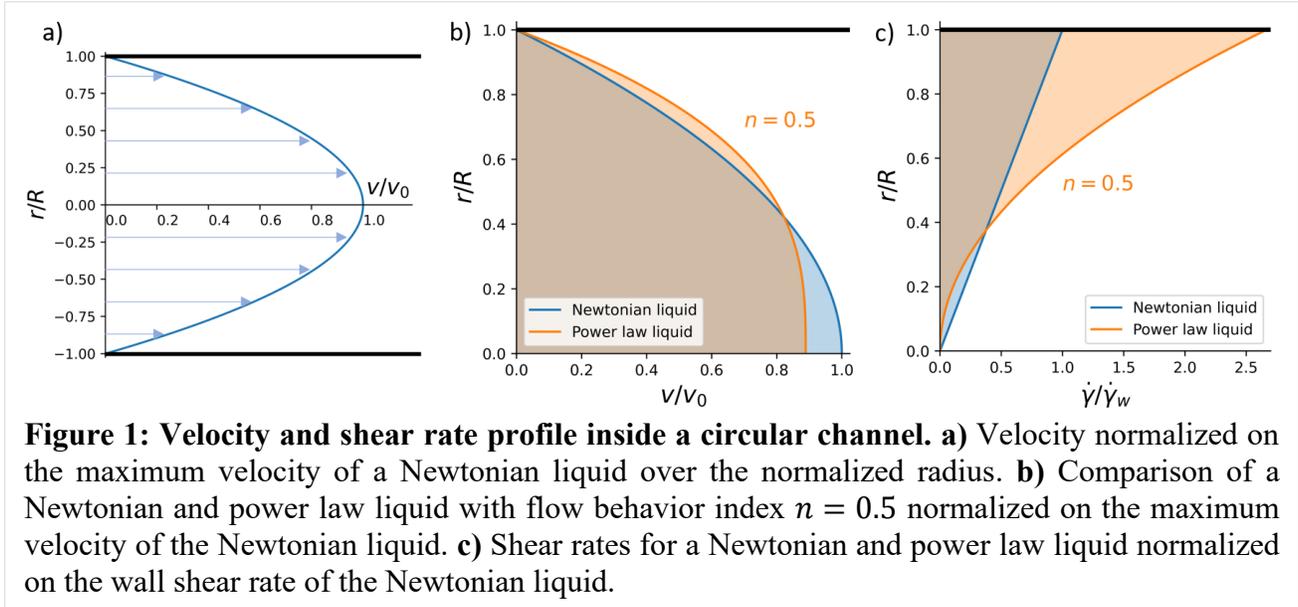

**Figure 1: Velocity and shear rate profile inside a circular channel. a)** Velocity normalized on the maximum velocity of a Newtonian liquid over the normalized radius. **b)** Comparison of a Newtonian and power law liquid with flow behavior index $n = 0.5$ normalized on the maximum velocity of the Newtonian liquid. **c)** Shear rates for a Newtonian and power law liquid normalized on the wall shear rate of the Newtonian liquid.

The resulting velocity profile is shown in figure 1. The velocity equations for flow in rectangular or square channels, like used for RT-DC, are more complicated to derive but the behavior will be similar.

From equation 4 one can derive the shear rates over the channel radius:

$$\dot{\gamma}(r) = \frac{\partial}{\partial r} v(r) = -\frac{v_0}{R} \cdot \frac{n+1}{n} \cdot \left(\frac{r}{R}\right)^{\frac{1}{n}}. \qquad (7)$$

The resulting shear rates for a Newtonian and power law liquid are shown in Figure 1c.

It becomes clear that the distribution of velocities over the channel cross section will lead to a distribution of shear rates and with that to a distribution of viscosities inside the channel for a shear thinning liquid. To define one *effective viscosity* at a given flow condition, Herold[2] proposed to use the viscosity based on the *wall shear rate* $\dot{\gamma}_w$. Wittwer et al.[3] recently verified in FEM simulations that using a shear thinning liquid or a Newtonian liquid with the viscosity based on the wall shear rate will lead to very similar cell deformations in RT-DC. For a given flow rate $Q$, Son derived an equation to determine the wall shear rate inside a rectangular channel for a power law liquid:[4]

$$\dot{\gamma}_w = \frac{8Q}{WH^2} \left[ b^*\left(\frac{H}{W}\right) + a^*\left(\frac{H}{W}\right) \cdot \frac{1}{n} \right] \qquad (8)$$

where $W$ is the channel width, $H$ is the channel height and $H < W$, $a^*$ and $b^*$ are geometrical factors dependent on the aspect ratio of the channel $H/W$ and were determined numerically (for a list of values, see Son[4]) and $n$ is the flow behavior index. By inserting equation 8 into equation 1, one can calculate the effective viscosity inside the channel. In the case of RT-DC channels, $H = W$ and $a^* = 0.2121$ and $b^* = 0.6771$. This leads to the following equation to calculate the effective viscosity in RT-DC:

$$\eta = K \cdot \left[\frac{8Q}{W^3}\left(0.6671 + \frac{0.2121}{n}\right)\right]^{n-1} \qquad (9)$$

## 2   Viscosity equations for MC-PBS solutions RT-DC

With the parameters given in Table 1 and equations 2, 3 and 9, it is possible to determine the viscosity of each MC-PBS solution for a given combination of channel size, flow rate and temperature. However, when closely inspecting the values given in Table 1 it can be seen that the errors on the values are quite large and that the parameters $\alpha$ and $\lambda$ show no clear dependence on the MC concentration. It is also expected that these parameters should be material constants of MC dissolved in PBS, within the concentration regime studied by Büyükurgancı et al. Furthermore, when calculating the viscosities of the 0.49% MC and 0.59% MC solutions, it can happen that under certain conditions the 0.49% solution has a higher viscosity than the 0.59% solution, which was never observed in rheometer measurements and is also not sensible.

Since the parameters $\alpha$ and $\lambda$ are not expected to change with concentration and can be considered constant within error margins, we propose to describe the viscosity functions with the averaged values $\bar{\alpha}$ and $\bar{\lambda}$ of the three concentrations. This gives the following values (mean $\pm$ SEM $\pm$ propagated error):

$$\bar{\alpha} = (2.23 \pm 0.07 \pm 0.37) \times 10^{-3} \text{ K}^{-1}$$

$$\bar{\lambda} = (3379.7 \pm 141.0 \pm 376.7) \text{ K}$$

The values for $\beta$ and $A$ have to be updated accordingly to best fit the data. The resulting curves are given in Figure 2. The resulting values for $\beta$ and $A$ are given in Table 2.

With the averaged parameters $\bar{\alpha}$ and $\bar{\lambda}$, the values from Table 2 and equations 2, 3 and 9 one can now reliably calculate the viscosity of the MC-PBS solutions in RT-DC experiments.

Table 2: Updated parameters $\beta$ and $A$ for the averaged values $\bar{\alpha}$ and $\bar{\lambda}$.

| Solution | $\beta$ | $A \times 10^{-6}$ [Pa s] |
|---|---|---|
| 0.49% MC-PBS | -0.0056 $\pm$ 0.0015 | 2.30 $\pm$ 0.05 |
| 0.59% MC-PBS | -0.0744 $\pm$ 0.0012 | 5.70 $\pm$ 0.07 |
| 0.83% MC-PBS | -0.1455 $\pm$ 0.0010 | 16.52 $\pm$ 0.30 |

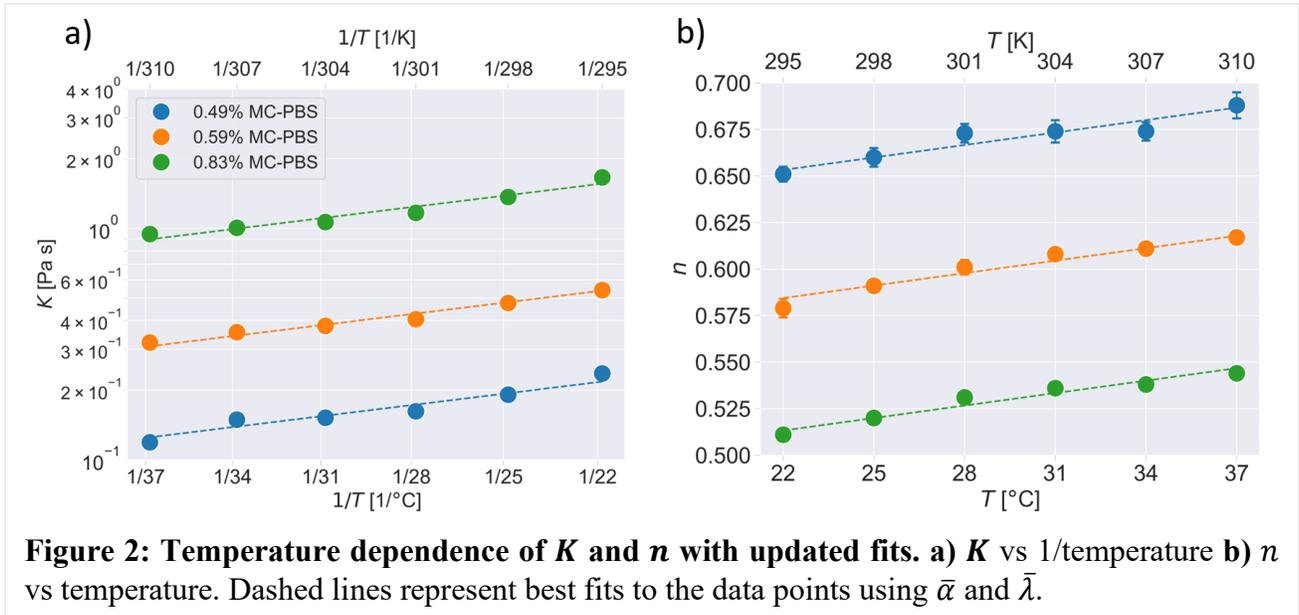

**Figure 2:** Temperature dependence of *K* and *n* with updated fits. a) *K* vs 1/temperature b) *n* vs temperature. Dashed lines represent best fits to the data points using $\bar{\alpha}$ and $\bar{\lambda}$.

## 3   Comparison of the new viscosity model with earlier results

Currently, the viscosity model used for the MC-PBS solutions is based on the work of Herold[2]. This model was derived based on measurements of the pressure drop and flow rate in a fixed channel geometry at a constant temperature. The temperature dependence was then measured separately in a falling ball viscometer in a range of 18 – 26 °C. The disadvantage of this method is that the shear rate of the measurement in the viscometer cannot be controlled. When the temperature changes, the viscosity of the medium will change and with that the speed at which the ball falls through the viscometer. This leads to measurements at higher shear rates for increasing temperatures. Thus, the temperature dependence of the shear thinning could not be determined with this approach.

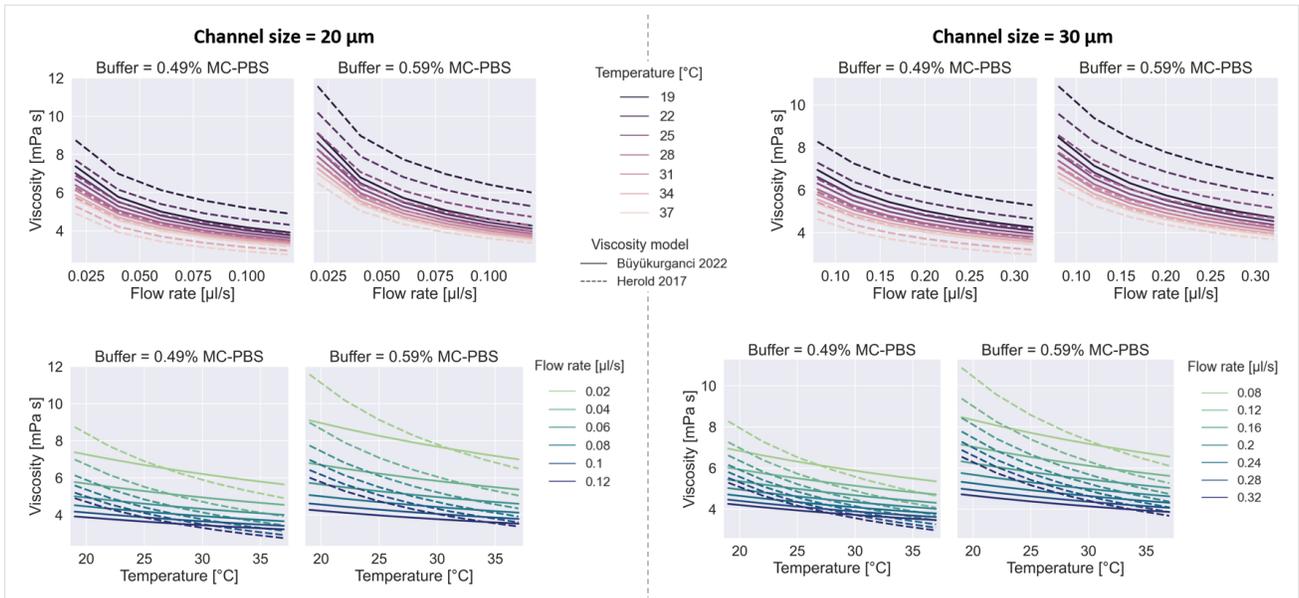

**Figure 3:** Comparison of viscosity curves from Büyükurgancı et al.[1] and Herold[2].

The viscosity curves from Büyükgancı et al. and Herold for typical channel sizes, flow rates and temperatures used in RT-DC experiments is shown in Figure 3. It can be seen that the temperature dependence in the Herold model is much stronger compared to Büyükgancı et al.

## 4 Summary and conclusions

Here, we provided a formalism to calculate the viscosity of three MC-PBS solutions, commonly used in RT-DC experiments for a given channel size, flow rate and temperature. The final equations have the following form:

**0.49% MC-PBS:**
$$\eta(W, Q, T) = 2.30 \cdot 10^{-6} \cdot \exp\left(\frac{3379.7\,[\text{K}]}{T}\right) \cdot \left[\frac{8Q}{W^3}\left(0.6671 + \frac{0.2121}{0.00223\frac{T}{[\text{K}]} - 0.0056}\right)\right]^{0.00223\frac{T}{[\text{K}]} - 1.0056} \text{Pa s}$$

**0.59% MC-PBS:**
$$\eta(W, Q, T) = 5.70 \cdot 10^{-6} \cdot \exp\left(\frac{3379.7\,[\text{K}]}{T}\right) \cdot \left[\frac{8Q}{W^3}\left(0.6671 + \frac{0.2121}{0.00223\frac{T}{[\text{K}]} - 0.0744}\right)\right]^{0.00223\frac{T}{[\text{K}]} - 1.0744} \text{Pa s}$$

**0.83% MC-PBS:**
$$\eta(W, Q, T) = 16.52 \cdot 10^{-6} \cdot \exp\left(\frac{3379.7\,[\text{K}]}{T}\right) \cdot \left[\frac{8Q}{W^3}\left(0.6671 + \frac{0.2121}{0.00223\frac{T}{[\text{K}]} - 0.1455}\right)\right]^{0.00223\frac{T}{[\text{K}]} - 1.1455} \text{Pa s}$$

The new viscosity equations will be made available for analysis of the Young's moduli to the RT-DC community through the python package *dclab*.[5]

The new viscosity equations can lead to rather large differences in the resulting viscosity compared to the model from Herold. This is possibly cause by an overestimation of the temperature influence in the Herold model because the temperature curve could not be determined for controlled shear rates. This will lead to changes in the resulting Young's modulus of objects in RT-DC based on the lookup tables (LUTs) introduced by Mokbel et al.[6] and Wittwer et al.[3] However, the LUTs are independent of viscosity model and the resulting Young's modulus scales linearly with the effective viscosity of the medium.[3] This will result in different magnitudes of the Young's modulus when analyzing with either of the two models but relative changes between two conditions will be preserved. But the comparison of Young's moduli computed from the Herold model should not be compared with those computed from the Büyükgancı model and we strongly discourage people to do so!

## 5 Acknowledgements

We want to thank Paul Müller for revising this manuscript and making the viscosity model available to the community through dclab.

# 6 Data availability

The analysis files for this manuscript will be available on gitlab: https://gitlab.gwdg.de/shear-rheology-of-methyl-cellulose-solutions/mc-viscosity-in-rtdc.

# 7 References

1 B. Büyükurgancı, S. K. Basu, M. Neuner, J. Guck, A. Wierschem and F. Reichel, Shear rheology of methyl cellulose based solutions for cell mechanical measurements at high shear rates, *Soft Matter*, DOI:10.1039/D2SM01515C.

2 C. Herold, Mapping of Deformation to Apparent Young's Modulus in Real-Time Deformability Cytometry, *arXiv*, 2017, 1704.00572.

3 L. D. Wittwer, F. Reichel, P. Müller, J. Guck and S. Aland, A New Hyperelastic Lookup Table for RT-DC, *arXiv*, 2022, 2208.12552.

4 Y. Son, Determination of shear viscosity and shear rate from pressure drop and flow rate relationship in a rectangular channel, *Polymer*, 2007, **48**, 632–637.

5 P. Müller and and others, dclab version 0.47.5: Python package for real-time deformability cytometry., https://github.com/DC-analysis/dclab.

6 M. Mokbel, D. Mokbel, A. Mietke, N. Träber, S. Girardo, O. Otto, J. Guck and S. Aland, Numerical Simulation of Real-Time Deformability Cytometry To Extract Cell Mechanical Properties, *ACS Biomater. Sci. Eng.*, 2017, **3**, 2962–2973.